\documentclass[aps,twocolumn,superscriptaddress,prl,amsmath,amssymb,10pt,aps]{revtex4-1} 

\usepackage[colorlinks=true,linkcolor=blue,citecolor=blue,urlcolor=blue]{hyperref}
\usepackage{graphicx}
\usepackage{latexsym}
\usepackage{amsmath}
\usepackage{wasysym}
\usepackage{amsthm}
\usepackage{amsbsy}
\usepackage{amssymb}
\usepackage{epstopdf} 
\usepackage{enumerate}
\usepackage{setspace} 
\usepackage{dcolumn} 
\usepackage{bm} 
\usepackage{slashed}
\usepackage{color}
\usepackage{youngtab}
\usepackage{float}
\usepackage{graphicx}
\usepackage{multirow}
\usepackage{array}
\usepackage{mathtools}
\usepackage{inputenc}
\usepackage{color}
\usepackage{makecell}

\begin{document}

\title{Electric Quantum Oscillation in Weyl Semimetals}

\author{Kyusung Hwang}
\affiliation{School of Physics, Korea Institute for Advanced Study, Seoul 02455, Korea}

\author{Woo-Ram Lee}
\affiliation{School of Physics, Korea Institute for Advanced Study, Seoul 02455, Korea}
\affiliation{Department of Physics, Virginia Tech, Blacksburg, Virginia 24061, USA}

\author{Kwon Park}
\affiliation{School of Physics, Korea Institute for Advanced Study, Seoul 02455, Korea}

\date{\today}
\begin{abstract}
Electronic transport in Weyl semimetals is quite extraordinary due to the topological property of the chiral anomaly generating the charge pumping between two distant Weyl nodes with opposite chiralities under parallel electric and magnetic fields.
Here, we develop a full nonequilibrium quantum transport theory of the chiral anomaly, based on the fact that the chiral charge pumping is essentially nothing but the Bloch oscillation. 
Specifically, by using the Keldysh nonequilibrium Green function method, it is shown that there is a rich structure in the chiral anomaly transport, including the negative magnetoresistance, the non-Ohmic behavior, the Esaki-Tsu peak, and finally the resonant oscillation of the DC electric current as a function of electric field, called the electric quantum oscillation. 
We argue that, going beyond the usual behavior of linear response, the non-Ohmic behavior observed in BiSb alloys can be regarded as a precursor to the occurrence of electric quantum oscillation, which is both topologically and energetically protected in Weyl semimetals.
\end{abstract}
\maketitle 

Among the greatest mysteries in physics is the asymmetry between matter and antimatter in our known universe. 
Baryogenesis  is the hypothesized physical process producing such an asymmetry, whose precise mechanism still remains elusive.
The chiral anomaly, also known as the Adler-Bell-Jackiw anomaly~\cite{Adler1969,Bell_Jackiw1969,Nielsen_Ninomiya1983}, is generally regarded as one of the most crucial elements in the mechanism of baryogenesis.

In this context, Weyl semimetals~\cite{Burkov2011,Son2013,Kim2013,Potter2014,Xu2015,Yang2015,Weng2015,Lv2015,Huang2015_NatCommun,Huang2015_PRX,Xiong2015,Li2016_NatCommun,Li2016_NatPhys,Shin2017,Armitage2018,Burkov2018} have been recently attracting intense attention by providing a concrete realization of the chiral anomaly in condensed matter systems, whose parameters can be tuned in tabletop experiments. 
This experimental tunability is highly useful to investigate various aspects of the chiral anomaly. 
A specific aspect of the chiral anomaly, which has attracted particularly intense attention, is the negative magnetoresistance (MR), i.e., the resistance decreases with stronger magnetic fields~\cite{Nielsen_Ninomiya1983,Son2013}.
While definitely important, however, the negative MR is ultimately a semiclassical signature of the chiral anomaly.

Here, we propose a full quantum signature of the chiral anomaly, which is fundamentally due to the quantization of the chiral charge pumping under strong electric fields.
A main breakthrough in this work is the realization that the chiral charge pumping is essentially nothing but the Bloch oscillation in the zeroth, or chiral Landau level (LL), 
which is quantized to generate robust Wannier-Stark ladder (WSL) eigenstates~\cite{Mendez1993,Raizen1997,Gluck2002} topologically protected by the chiral anomaly. 
Albeit somewhat less, robust WSL eigenstates can be also formed in nonchiral LLs due to the energetic protection of the Bloch oscillation in Weyl semimetals.

The formation of WSL eigenstates reveals an intriguing similarity between electricity and magnetism.
The quantized cyclotron motion of electrons under strong magnetic fields gives rise to well-known magnetic quantum oscillations~\cite{Shoenberg_Book} such as de Haas-van Alphen, Shubnikov-de Haas, and eventually the quantum Hall effects.
Similarly, the quantized Bloch oscillation of electrons under strong electric fields can give rise to an electric-field-induced oscillation of the DC electric current, which we call the electric quantum oscillation (EQO).

Actually, the EQO brings out one of the most fundamental differences between electricity and magnetism.
That is, electric fields inevitably cause nonequilibrium, while magnetic fields do not, no matter how strong.
This difference raises a pressing question.
What is the nonequilibrium steady state induced by strong electric fields?

In this work, we develop a full nonequilibrium quantum transport theory of the chiral anomaly by using the Keldysh nonequilibrium Green function formalism~\cite{Haug_Jauho_Book} in conjunction with the Lindblad quantum master equation~\cite{Lee2014}.
As a result, it is shown that there is a rich structure in the chiral anomaly transport, including the negative MR, the non-Ohmic behavior, the Esaki-Tsu peak, and finally the EQO. 
Being the incipient nonlinear behavior characterizing the chiral anomaly transport, the non-Ohmic behavior observed in BiSb alloys~\cite{Shin2017} can be regarded as a precursor to the occurrence of EQO, which can serve as the unmistakable quantum signature of the chiral anomaly in Weyl semimetals.
We emphasize that the chiral anomaly provides a unique environment for the realization of WSL eigenstates in natural materials, which has been so far impossible except for synthetic systems such as semiconductor superlattices~\cite{Mendez1988,Voisin1988} and optical lattices~\cite{Wilkinson1996,Dahan1996}.

In the perspective of application, this work lays a groundwork to expand the frontier of nonequilibrium quantum transport and realize novel nonlinear electronic devices by combining strong-field phenomena~\cite{Kruchinin2018} with topological matter.
It is interesting to mention that strong-field phenomena have been also investigated in combination with various many-body correlation effects such as Mott transition~\cite{Freericks2006,Tsuji2008,Eckstein2010,Amaricci2012,Aron2012,Lee2014,Mazza2016,Diener2018} and many-body localization~\cite{Schulz2019,Nieuwenburg2019}.

\begin{figure*}
\centering
\includegraphics[width=\textwidth]{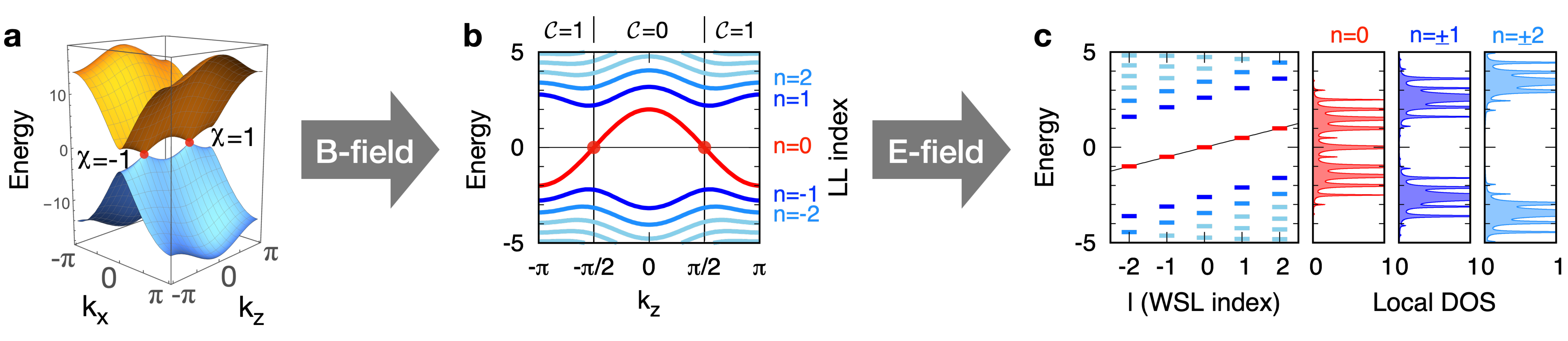}
\caption{{\bf Landau-Stark quantization.}
({\bf a}) Energy spectrum of a minimal tight-binding model for Weyl semimetals without external fields.
$\chi$ denotes the chirality of each Weyl node.
({\bf b}) Energy spectrum under a magnetic field applied in the $z$ direction, showing the formation of Landau level (LL) eigenstates with the energy eigenvalue of $\epsilon_n(k_z)$. 
With each LL labeled by the LL index $n$, the $n=0$, or chiral LL is plotted in red, while all other nonchiral LLs are in dark/light blue.
${\cal C}$ denotes the Chern number of the 2D $k_z$ slices of the Brillouin zone before the application of magnetic fields.
({\bf c}) Energy spectrum under parallel electric and magnetic fields applied in the $z$ direction, showing the formation of Wannier-Stark ladder (WSL) eigenstates in each LL, called Landau-Stark eigenstates, with the energy eigenvalue of $\epsilon_{nl}=\bar{\epsilon}_n+l\Omega$.
The Landau-Stark quantization manifests itself as a series of discrete peaks in the local density of states (DOS) shown in right panels indicating individual contributions from various LLs.
Here, the cyclotron and Bloch oscillation frequencies are set as $\omega_c=0.3$ and $\Omega=0.5$ in units of $t_3$, respectively.
}
\label{fig:Landau-Stark}
\end{figure*}

\noindent{\bf Landau-Stark quantization}

\noindent To perform a concrete analysis of the chiral anomaly transport in the full quantum level, we consider a minimal tight-binding model for Weyl semimetals~\cite{Yang2011,Delplace2012}:
\begin{equation}
H({\bf k}) = \sum_{i=x,y,z} h_i({\bf k}) \sigma_i,
\label{eq:H_k}
\end{equation}
where $h_x({\bf k})=2t_1\sin k_x$, $h_y({\bf k})=-2t_1\sin k_y$, and $h_z({\bf k})=2t_2 [ 2-\cos(k_x-k_y)-\cos(k_x+k_y)]
+2t_3\cos k_z$ with $t_1, t_2, t_3$ being hopping amplitudes and $\sigma_x, \sigma_y, \sigma_z$ being Pauli matrices.
With the time-reversal symmetry broken, this model hosts a single pair of Weyl nodes at ${\bf k}=(0,0,\pm\pi/2)$ with zero energy. 
See Fig.~\ref{fig:Landau-Stark}~{\bf a} for the illustration of the energy spectrum at $t_1=4$, $t_2=2$, and $t_3=1$, which are to be used as hopping amplitudes throughout this work.
Note that all momenta are denoted in units of corresponding inverse lattice constants.
Also, unless specified otherwise, we set $\hbar=c=1$ for simplicity.
As elaborated later, in this work, we focus on half filling by setting the chemical potential appropriately.

Let us first investigate what happens to the energy spectrum of the model Hamiltonian with the application of magnetic fields in the $z$ direction (${\bf B} =B \hat{z}$ with $B>0$).
Actually, the energy spectrum would develop a highly complicated fractal structure known as Hofstadter's butterfly, if magnetic fields are directly applied to the lattice model.
To avoid this complication, we take the continuum limit of the model Hamiltonian within the $x$-$y$ plane by replacing $\sin{k_i}$ by $k_i$ and $\cos{k_i}$ by $1-k_i^2/2$ for $i=x, y$ in Eq.~\eqref{eq:H_k}, while maintaining the full $k_z$ dispersion.

The application of magnetic fields can be implemented via minimal coupling, i.e., ${\bf k} \rightarrow -i\nabla+e{\bf A}_{\rm Landau}$ with $-e$ being the charge of electron and ${\bf A}_{\rm Landau}$ being the Landau-gauge vector potential. 
Consequently, the model Hamiltonian generates the following energy eigenvalues under magnetic fields:
\begin{equation}
\epsilon_n(k_z)={\rm sgn}(n)\sqrt{ (2t_3 \cos{k_z} + |n|\omega_c)^2 + 2|n|\omega_c t_1^2/t_2},
\label{eq:LL_energy_nonchiral}
\end{equation}
where $n$ is a nonzero integer, and $\omega_c=4t_2 a_x a_y /l_B^2$ is the cyclotron frequency with $a_x$ and $a_y$ being the lattice constants in the $x$ and $y$ directions, respectively, and $l_B =1/\sqrt{eB}$ being the magnetic length.
For simplicity, the zero-point energy $\omega_c/2$ is subtracted from the energy eigenvalues to define $\epsilon_n(k_z)$. 
Note that the energy eigenmodes are composed of the usual LL eigenstates, which are entirely dispersionless within the $x$-$y$ plane, while dispersive in the $z$ direction.

Now, an interesting thing happens if one tries to set $n=0$ in Eq.~\eqref{eq:LL_energy_nonchiral}.
With the sign of zero undefined, there could be two distinct energy eigenmodes corresponding to $\pm2t_3 |\cos{k_z}|$.
In reality, however, there exists only a single energy eigenmode called the chiral LL with the energy eigenvalue of $\epsilon_{0}(k_z)=2t_3\cos{k_z}$.
Note that this singleness of the chiral LL is a unique topological property of Weyl semimetals. 
See {\bf Methods} for details.
Also, see Fig.~\ref{fig:Landau-Stark}~{\bf b} for the illustration of chiral versus nonchiral LLs.

With the application of electric fields, each LL can be further quantized into a series of WSL eigenstates.
Usually, the formation of WSL eigenstates requires well-separated energy bands so that the Bloch oscillation can complete one full cycle without being interrupted by the Landau-Zener transition~\cite{Gluck2002}, which is unfortunately difficult to achieve in natural materials. 
Fortunately, in Weyl semimetals, there is a nice protection of the Bloch oscillation due to the aforementioned singleness of the chiral LL. 
Specifically, when electric fields are applied in the $z$ direction parallel to magnetic fields (${\bf E} =E\hat{z}$), there is absolutely no Landau-Zener transition between different LLs, unless they have the same $|n|$.
This means that the Bloch oscillation in the chiral LL is completely immune from the Landau-Zener transition.
While allowed, the Landau-Zener transition is also energetically suppressed between nonchiral LLs with the same $|n|$, whose energies can be well-separated across the zero-point energy.
Consequently, under strong magnetic fields, it is safe to assume that each LL is quantized into its own individual series of WSL eigenstates.

Also, being so-called extended states, LL eigenstates are generally known to be rather robust against disorder~\cite{DasSarma_Pinczuk_Book}. 
This means that the $k_z$ dispersion of LL eigenstates and consequently the formation of WSL eigenstates can be also robust against disorder to certain extents.

Technically, the application of electric fields can be implemented in terms of either static scalar or temporal vector potential gauge.
In the static scalar potential gauge, the model Hamiltonian can be written in terms of the Stark Hamiltonian for each individual LL:
\begin{equation}
H_{{\rm Stark},n}= \epsilon_{n}(k_z)+i\Omega\left[\frac{\partial}{\partial k_z} +{\cal A}_{n}(k_z)\right],
\label{eq:H_Stark}
\end{equation}
where $\Omega=eEa_z$ is the Bloch oscillation frequency with $a_z$ being the lattice constant in the $z$ direction.
Here, ${\cal A}_{n}(k_z)$ is the Berry connection of the $n$-th LL, which turns out to be zero regardless of $n$ in our minimal model.
The Stark Hamiltonian can be diagonalized via WSL eigenstates in each LL, called Landau-Stark eigenstates, i.e., 
$H_{{\rm Stark},n}\phi_{nl}(k_z)=\epsilon_{nl}\phi_{nl}(k_z)$ with
\begin{equation}
\phi_{nl}(k_z) = e^{-\frac{i}{\Omega} \int_0^{k_z} d \kappa [\epsilon_{nl} - \epsilon_{n}(\kappa)]} 
\label{eq:Landau-Stark_eigenstates}
\end{equation}
and $\epsilon_{nl}= \bar{\epsilon}_n +l\Omega$, where $\bar{\epsilon}_{n}=\int_{-\pi}^{\pi}\frac{\textup{d}k_z}{2\pi}\epsilon_{n}(k_z)$ is the mean energy of the $n$-th LL and $l$ is the WSL index. 
It is important to note that WSL eigenstates are full quantum solutions of the Stark Hamiltonian, while also obtained as semiclassical solutions via the Bohr-Sommerfeld quantization~\cite{Xiao2010}.
See Fig.~\ref{fig:Landau-Stark}~{\bf c} for the illustration of Landau-Stark energy levels, accompanied by the local density of states (DOS), whose details are given in {\bf Methods}.

\noindent{\bf Nonequilibrium quantum transport}

\noindent Being standing waves, WSL eigenstates cannot generate any nonzero net DC electric currents, unless there is impurity scattering, which causes the transition between different WSL eigenstates.
Here, we develop a full nonequilibrium quantum transport theory of the chiral anomaly by treating the process of impurity scattering via the Keldysh nonequilibrium Green function formalism~\cite{Haug_Jauho_Book}. 
Specifically, our nonequilibrium quantum transport theory is composed of three steps.

\noindent{\bf Temporal vector potential gauge.}
The first step is to change the gauge and implement the application of electric fields via the temporal vector potential ${\bf A}_{\rm Stark}=-Et\hat{z}$ with $t$ being time, in which case the total vector potential is given as ${\bf A}={\bf A}_{\rm Landau}+{\bf A}_{\rm Stark}$.
This particular choice of gauge is made to preserve the spatial translation symmetry so that impurity scattering can be treated via the usual method of self-consistent Born approximation (SCBA). 

In the temporal vector potential gauge, the model Hamiltonian can be written as follows:
\begin{equation}
H_n(t)= \epsilon_{n}(k_z-\Omega t),
\label{eq:H_time}
\end{equation}
which is periodic in time with the period of $2\pi/\Omega$. 
Such a time-dependent Hamiltonian can be analyzed by using the Keldysh nonequilibrium Green function method with nonequilibrium Green functions conveniently represented in the Floquet matrix form~\cite{Tsuji2008}.
It is worthwhile to mention that WSL eigenstates in the static scalar potential gauge are manifested as Floquet modes in the temporal vector potential gauge.

\begin{figure}
\centering
\includegraphics[width=0.8\linewidth]{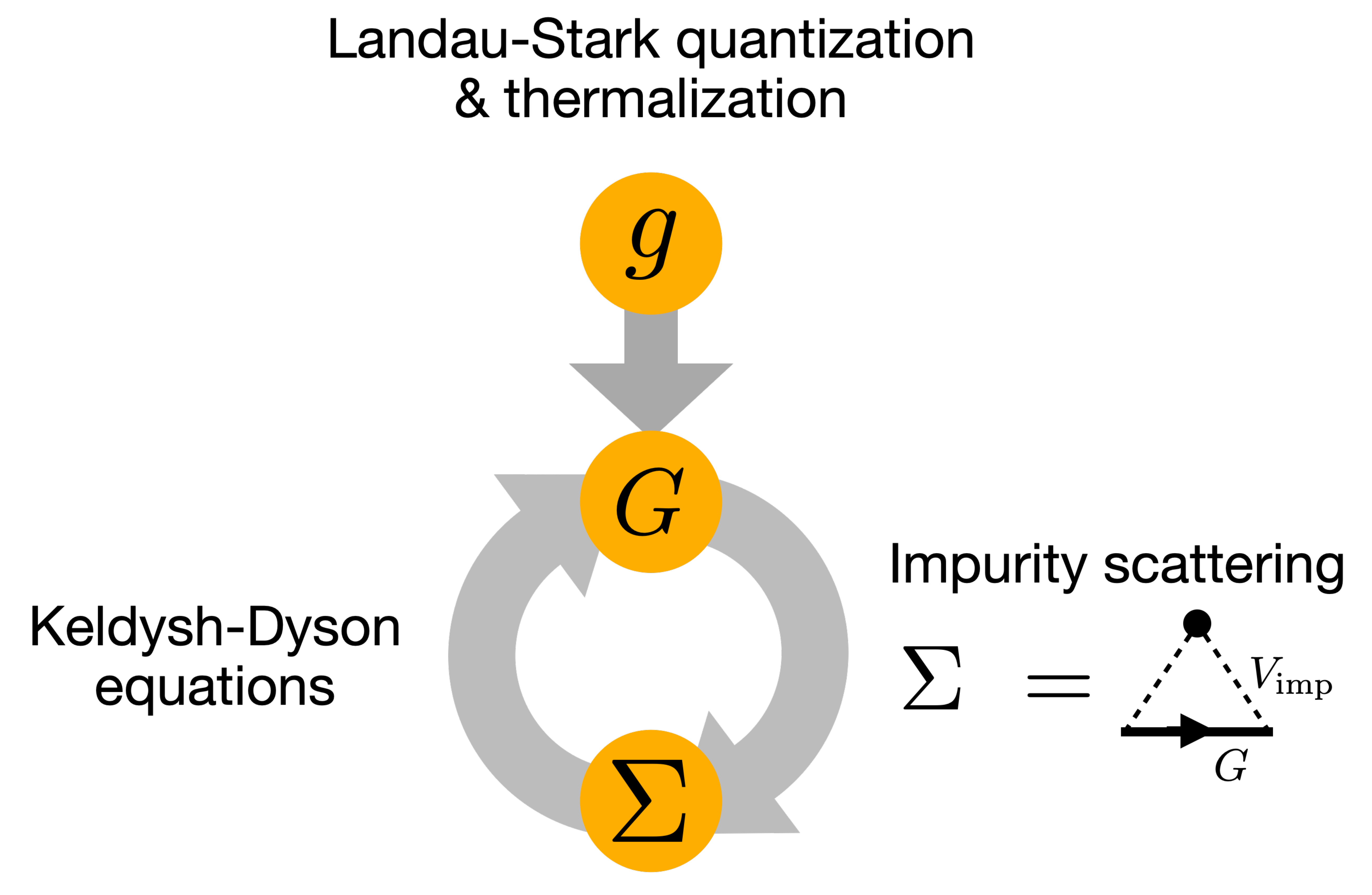}
\caption{{\bf Schematic diagram of the Keldysh-Dyson self-consistency loop.}
Our nonequilibrium quantum transport theory is based on the Keldysh-Dyson self-consistency loop comprising three parts;
(i) the full Green function, $G$, is obtained by solving the Keldysh-Dyson equations with $\Sigma$ being the yet-to-be-determined self-energy,
(ii) $\Sigma$ is then related with $G$ via self-consistent Born approximation for impurity scattering, and
(iii) the self-consistency loop is completed once the noninteracting Green function, $g$, is fixed in terms of Landau-Stark eigenstates.
Crucially, the noninteracting lesser Green function, $g^<$, is constructed so that Landau-Stark eigenstates are appropriately thermalized according to the WSL-wise thermalization scheme.
}
\label{fig:self-consistency_loop}
\end{figure}

\noindent{\bf Keldysh-Dyson self-consistency loop.}
The second step is to set up the Keldysh-Dyson self-consistency loop to capture the process of impurity scattering via SCBA.
See Fig.~\ref{fig:self-consistency_loop} for the schematic diagram.
Technically, the full Green functions can be obtained by self-consistently solving the Keldysh-Dyson equations~\cite{Haug_Jauho_Book}:
\begin{gather}
[G^r]^{-1} = [g^r]^{-1}-\Sigma^r,
\label{eq:Dyson}
\\
G^< = G^r \{ [g^r]^{-1} g^< [g^a]^{-1} + \Sigma^< \} G^a,
\label{eq:Keldysh}
\end{gather}
where $G^r$ ($g^r$) and $G^<$  ($g^<$) are the full (noninteracting) retarded and lesser Green functions, which contain the information about the DOS and occupation, respectively.
The advanced Green functions, $G^a$ and $g^a$, are related with the retarded counterparts via complex conjugation. 
Meanwhile, $\Sigma^{r}$ and $\Sigma^{<}$ are the retarded and lesser self-energies, respectively, induced by impurity scattering.
In the above expressions, we drop all the subscripts (LL and Floquet indices) and arguments ($k_z$ and $\omega$) for simplicity.

Importantly, the self-energies are related to the full Green functions via SCBA ~\cite{Bruus_Flensberg_Book}:
\begin{equation}
\Sigma^{r,<}(\omega)=V_{\rm imp}^2 {\cal D}  \int_{-\pi}^{\pi} d k_z \sum_{n} G_{n}^{r,<} (k_z,\omega),
\label{eq:SCBA}
\end{equation}
where $V_{\rm imp}$ is the strength of the on-site Coulomb interaction between electron and impurity.
The factor ${\cal D}=\omega_c/8\pi t_2$ comes from the degeneracy of each LL.

Now, the Keldysh-Dyson self-consistency loop is completed once the noninteracting Green functions, $g^r$ and $g^<$, are fixed.
The noninteracting retarded Green function, $g^r$, is given in the Floquet matrix form as follows~\cite{Tsuji2008,Lee2014}:
\begin{equation}
[g_{n}^r(k_z,\omega)]_{pq}=
e^{ik_z(p-q)} \sum_{j} 
\varphi_{np}(j) {\cal G}_n^r(\omega+j\Omega) \varphi_{nq}^*(j),
\label{eq:Floquet_g_retarded}
\end{equation}
where $-\Omega/2 \leq \omega \leq \Omega/2$, $p$ and $q$ are Floquet indices,
$\varphi_{nl}(j)=\int_{-\pi}^{\pi} \frac{dk_z}{2\pi} e^{ik_z j} \phi_{nl}(k_z)$ is the Fourier transform of Landau-Stark eigenstates in Eq.~\eqref{eq:Landau-Stark_eigenstates}, and 
${\cal G}_n^r(\varepsilon=\omega+j\Omega)=1/(\varepsilon-\bar{\epsilon}_n+i\eta)$ is the reduced retarded Green function of Landau-Stark eigenstates. 
Note that the spectral information of Landau-Stark eigenstates is encoded via $\varphi_{nl}$ and ${\cal G}_n^r$ in $g_n^r$.
See {\bf Methods} for details. 

At this point, it is important to incorporate the broadening of Landau-Stark energy levels, which can be caused by any additional processes of inelastic scattering beyond SCBA.
We implement such inelastic level broadening by setting $\eta=\Gamma/2$ in ${\cal G}_n^r$ with $\Gamma$ being small, but finite.
Specifically, we set $\Gamma/t_3=0.05$ throughout this work.
Note that the DC electric current would be net zero in the presence of $\Gamma$ alone~\cite{Lee2014}. 
Nonzero net DC electric currents can be only generated by the intricate interplay of both elastic and inelastic scattering.

The noninteracting lesser Green function, $g^<$, is given in the Floquet matrix form as follows:
\begin{equation}
[g_{n}^<(k_z,\omega)]_{pq}=
e^{ik_z(p-q)} \sum_{j} 
\varphi_{np}(j) {\cal G}_n^<(\omega+j\Omega) \varphi_{nq}^*(j),
\label{eq:Floquet_g_lesser}
\end{equation}
where the reduced lesser Green function of Landau-Stark eigenstates, ${\cal G}_n^<(\varepsilon=\omega+j\Omega)$, is obtained via the WSL-wise thermalization scheme, which is in turn derived as a solution of the Lindblad quantum master equation~\cite{Lee2014}.
Specifically, each WSL eigenstate is individually thermalized according to the standard fluctuation-dissipation relation:
\begin{equation}
{\cal G}_n^<(\varepsilon) = [{\cal G}_n^a(\varepsilon)-{\cal G}_n^r(\varepsilon)] f_{\rm FD}(\varepsilon),
\label{eq:fluc_diss}
\end{equation}
where ${\cal G}_n^a(\varepsilon)={\cal G}_n^{r*}(\varepsilon)$ and $f_{\rm FD}(\varepsilon)=1/(e^{(\varepsilon-\mu)/k_B T}+1)$ is the usual Fermi-Dirac distribution function with the chemical potential $\mu$ set to be zero for half filling.
See {\bf Methods} for details. 

\noindent{\bf DC electric current density.}
The third and final step is to compute the DC electric current density, $J_{\rm DC}$, from the full lesser Green function obtained as a converged solution of the Keldysh-Dyson self-consistency loop~\cite{Lee2014}:
\begin{equation}
J_{\rm DC} =
e {\cal D}
\int_{-\Omega/2}^{\Omega/2} \frac{d\omega}{2\pi}
\sum_{n,p,q}
p \bar{\epsilon}_{n}(p)
\left[ G_{n}^< (k_z=0,\omega) \right]_{p+q,q},
\label{eq:J_DC}
\end{equation}
where $\bar{\epsilon}_{n}(p)=\int_{-\pi}^{\pi} \frac{dk_z}{2\pi}e^{i k_z p}\epsilon_{n}(k_z)$.
It is important to note that Eq.~\eqref{eq:J_DC} itself is an exact expression of $J_{\rm DC}$, which means that $J_{\rm DC}$ is accurate at arbitrary strengths of electric and magnetic fields if the full lesser Green function is so.
See {\bf Methods} for details. 

In the following sections, we present the numerical results of $J_{\rm DC}$ as a function of various parameters. 
Unless specified otherwise, all parameters with the energy unit (such as $\Omega$, $\omega_c$, $V_{\rm imp}$, $\Gamma$, and so on) are denoted in units of $t_3$ throughout this work.
Particularly, we set $k_B T/t_3=0.001$ in this work.
Also, LL indices are summed up to $|n|=6$, which is necessary for the range of magnetic fields studied in this work, except for the ultra-quantum limit of strong magnetic fields, where it is sufficient to consider only the chiral LL.
Meanwhile, the number of summed Floquet indices is chosen adaptively to ensure that $J_{\rm DC}$ is well converged at each given $\Omega$.
See {\bf Methods} for details. 

\begin{figure*}
\centering
\includegraphics[width=0.8\linewidth]{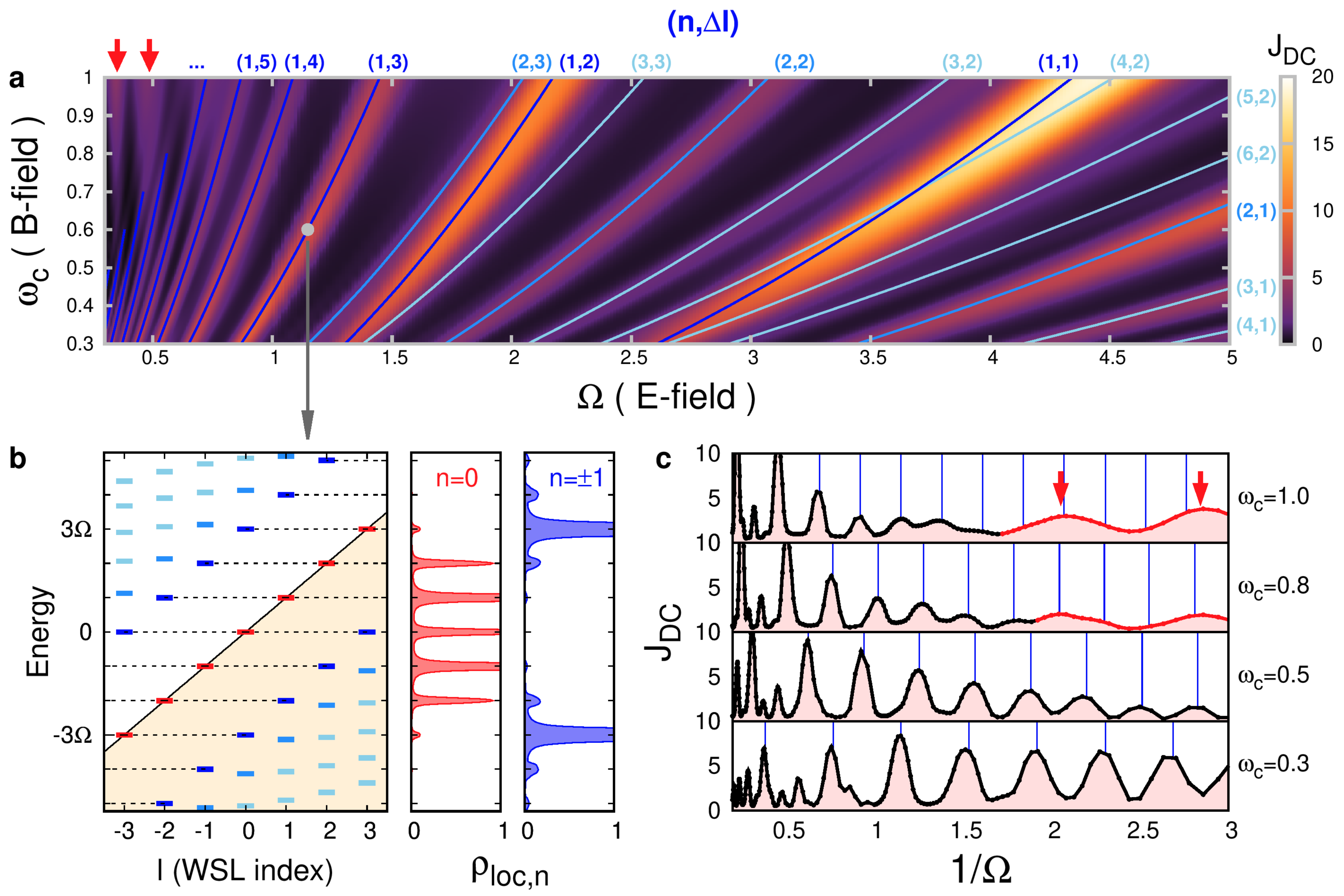}
\caption{{\bf Electric quantum oscillation via the general Landau-Stark resonance.}
({\bf a}) Color map of the DC electric current density $J_{\rm DC}$ as a function of the Bloch oscillation frequency $\Omega$ ($\propto E$) and the cyclotron frequency $\omega_c$ ($\propto B$).
Note that $J_{\rm DC}$ exhibits a complicated, yet highly organized series of resonant peaks, whose trajectories are accurately described by the general Landau-Stark resonance condition, $\bar{\epsilon}_n-\bar{\epsilon}_0=\Delta l \cdot \Omega$, for various cases of $(n,\Delta l)$.
The strongest resonant peaks are obtained along the trajectories of $(n=1,\Delta l=1, 2, 3, \cdots)$ plotted in dark blue, followed by progressively weaker resonant peaks along those of $n>1$ plotted in light blue. 
({\bf b}) Illustrated mechanism of the general Landau-Stark resonance.
Here, we take the case of $(n=1,\Delta l=3)$ as an example, marked by the grey dot in {\bf a}, where the Landau-Stark energy levels coming from the $n=1$ LL (dark blue) are perfectly aligned with those from the $n=0$, or chiral LL (red).
({\bf c}) Periodicity of $J_{\rm DC}$ as a function of $1/\Omega$ for various given $\omega_c$.
Generally, the resonant peaks are equally spaced as a function of $1/\Omega$ with the period of $1/(\bar{\epsilon}_n-\bar{\epsilon}_0)$, as shown by the blue vertical lines indicating the locations of the strongest resonant peaks at $n=1$.
There is an exception to this rule, as indicated by the red arrows here and also in {\bf a}, revealing the existence of a different type of the electric quantum oscillation.
Here, we set $V_{\rm imp}=1$ and $\Gamma=0.05$ in units of $t_3$.
Finally, $J_{\textup{DC}}$ is denoted in units of $10^{-5} e t_3/\hbar a_z^2$.
}
\label{fig:EQO_general}
\end{figure*}

\noindent{\bf Results}

\noindent{\bf Electric quantum oscillation via the general Landau-Stark resonance.}
The DC electric current can oscillate via two different mechanisms.
In this section, we first discuss the resonance between various Landau-Stark eigenstates with different LL indices, called the general Landau-Stark resonance.

Fig.~\ref{fig:EQO_general}~{\bf a} shows that, in a general regime of electric and magnetic fields, $J_{\rm DC}$ oscillates as a function of both $\Omega$ and $\omega_c$, exhibiting a complicated, yet highly organized series of resonant peaks. 
Physically, the resonant behavior of $J_{\rm DC}$ can be well understood in terms of the tunneling formula between adjacent sites~\cite{Lee2014}:
\begin{equation}
J_{\rm tun} \propto \int_{-\infty}^{\infty} d\varepsilon \rho_{\rm loc}(\varepsilon) \rho_{\rm loc}(\varepsilon+\Omega)
[f_{\rm loc}(\varepsilon)-f_{\rm loc}(\varepsilon+\Omega)],
\label{eq:J_tun}
\end{equation}
where $\rho_{\rm loc}$ and $f_{\rm loc}$ are the local DOS and distribution function, respectively. 
Specifically, $\rho_{\rm loc}$ is given as the sum of individual contributions from various LLs, i.e., $\rho_{\rm loc}={\cal D} \sum_n \rho_{{\rm loc},n}$ with
\begin{equation}
\rho_{{\rm loc},n}(\varepsilon=\omega+p\Omega) = -\frac{1}{\pi} \int_{-\pi}^{\pi} \frac{dk_z}{2\pi} {\rm Im}[G_n^r(k_z,\omega)]_{pp},
\label{eq:rho_loc}
\end{equation}
where $\varepsilon$ can cover the entire range of frequency by changing the Floquet index $p$ while $\omega \in [-\Omega/2,\Omega/2]$.
Meanwhile, $f_{\rm loc}$ can be computed via $f_{\rm loc} = N_{\rm loc}/\rho_{\rm loc}$ with the local occupation number $N_{\rm loc}$ given by $N_{\rm loc}={\cal D} \sum_n N_{{\rm loc},n}$, where
\begin{equation}
N_{{\rm loc},n}(\varepsilon=\omega+p\Omega) = \frac{1}{2\pi} \int_{-\pi}^{\pi} \frac{dk_z}{2\pi} {\rm Im}[G_n^<(k_z,\omega)]_{pp}.
\label{eq:N_loc}
\end{equation}
Note that Eq.~\eqref{eq:J_tun} can be formally derived from Eq.~\eqref{eq:J_DC} in the limit of strong electric fields, where WSL eigenstates form well-localized wave packets~\cite{Lee2014}.

According to the tunneling formula, the DC electric current can be maximized if there is a large overlap between $\rho_{\rm loc}(\varepsilon)$ and $\rho_{\rm loc}(\varepsilon+\Omega)$. 
Considering that $\rho_{\rm loc}(\varepsilon)$ is composed of periodic peaks due to the Landau-Stark quantization, this means that the DC electric current can be maximized along the trajectories in the $\Omega$-vs-$\omega_c$ parameter space, satisfying the following condition of the general Landau-Stark resonance:
\begin{equation}
\bar{\epsilon}_n-\bar{\epsilon}_0=\Delta l \cdot \Omega, 
\label{eq:resonance_condition}
\end{equation}
where $\Delta l$ is an integer.
See Fig.~\ref{fig:EQO_general}~{\bf b} for the illustrated mechanism of the general Landau-Stark resonance. 
As seen from Fig.~\ref{fig:EQO_general}~{\bf a}, the general Landau-Stark resonance condition describes the trajectories of maximized $J_{\rm DC}$ quite accurately. 
Note that a similar resonance phenomenon has been observed in the transport experiment of semiconductor superlattices under parallel electric and magnetic fields~\cite{Canali1996}.

Finally, to clearly show the periodicity of EQO, it is beneficial to plot $J_{\rm DC}$ as a function of $1/\Omega$ for various given $\omega_c$.
Fig.~\ref{fig:EQO_general}~{\bf c} shows that the resonant peaks are equally spaced as a function of $1/\Omega$ with the period of $1/(\bar{\epsilon}_n-\bar{\epsilon}_0)$, which is strongly reminiscent of the similar behavior in magnetic quantum oscillation.  
Actually, the low-electric-field data at $\omega_c=0.8$ and $1$ (red curves) reveals that, under strong magnetic fields, there is a new type of the EQO with different periodicity, which is shown below to be induced by a form of the self-resonance entirely within the chiral LL, called the chiral resonance.

\begin{figure}
\centering
\includegraphics[width=0.9\linewidth]{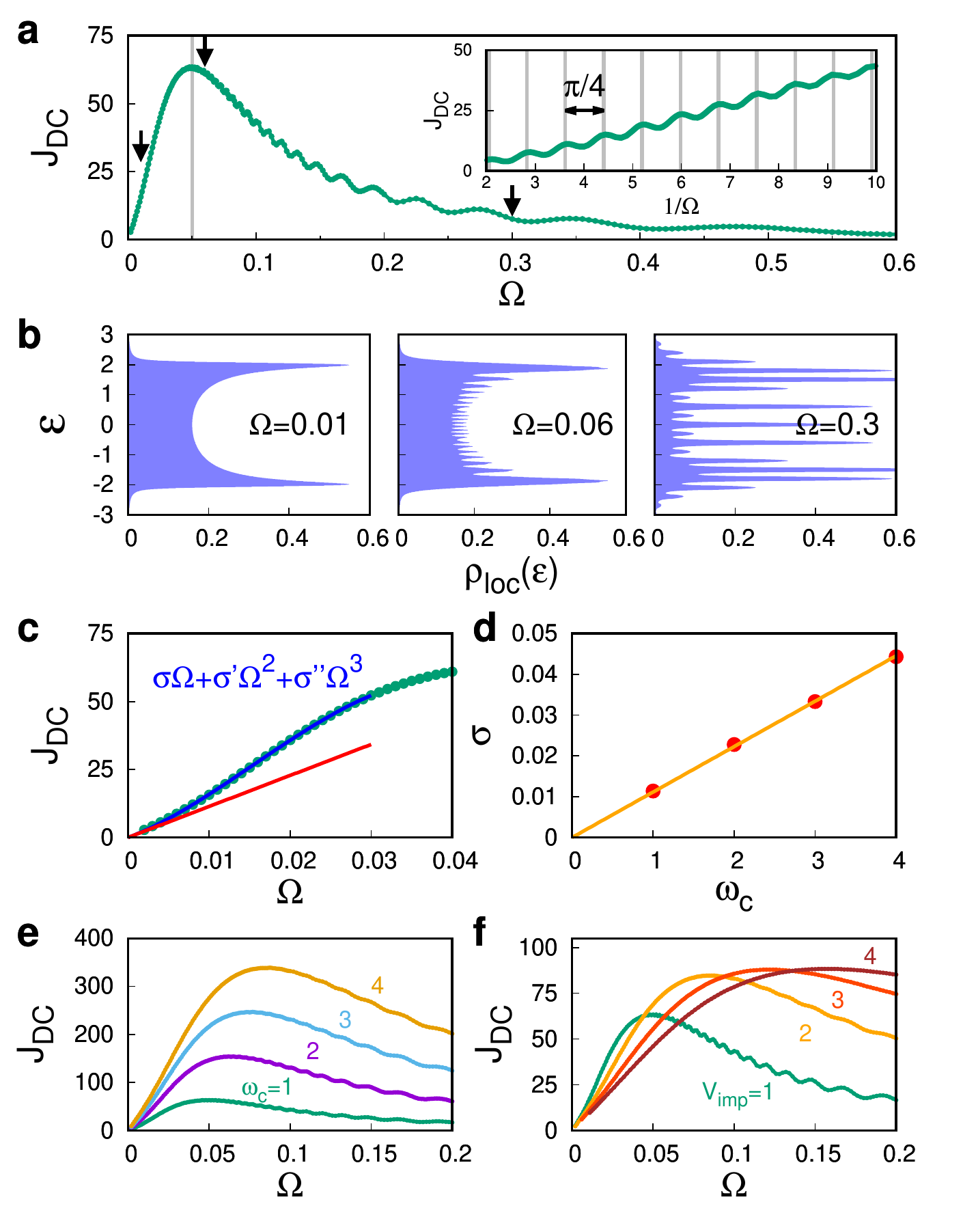}
\caption{{\bf Electric quantum oscillation via the chiral resonance.}
({\bf a}) DC electric current density $J_{\rm DC}$ as a function of $\Omega$ at $\omega_c=1$ with $V_{\rm imp}=1$, covering four distinct regimes of the chiral anomaly transport; (i) negative magnetoresistance, (ii) non-Ohmic behavior, (iii) Esaki-Tsu peak (indicated by the grey vertical line), and (iv)  electric quantum oscillation (EQO). 
The inset highlights the periodicity of the EQO via the chiral resonance as a function of $1/\Omega$, which is simply $\pi/4$ in units of $1/t_3$, being entirely independent of magnetic fields.
({\bf b}) Local DOS at $\Omega=0.01$, $0.06$, and $0.3$, indicated by the black arrows in {\bf a}. 
({\bf c}) Polynomial fitting of $J_{\rm DC}$ at weak and weak-to-intermediate electric fields, showing the usual linear Drude conductivity (red straight line) and the non-Ohmic behavior (blue curve), respectively.
({\bf d}) Magnetic-field dependence of the linear Drude conductivity, $\sigma$, showing the behavior of negative magnetoresistance.
({\bf e}) $J_{\rm DC}$ as a function of $\Omega$ for various given $\omega_c$, showing the overall increase of $J_{\rm DC}$ with stronger magnetic fields.
({\bf f}) $J_{\rm DC}$ as a function of $\Omega$ for various given $V_{\rm imp}$, showing the Drude behavior, i.e., $\sigma$ decreases with stronger impurity scattering.
As in Fig.~\ref{fig:EQO_general}, $J_{\textup{DC}}$ is denoted in units of $10^{-5} e t_3/\hbar a_z^2$.
}
\label{fig:EQO_chiral}
\end{figure}

\noindent{\bf Electric quantum oscillation via the chiral resonance.}
The general Landau-Stark resonance condition can be trivially satisfied with $n=0$ and $\Delta l=0$.
If so, na\"{i}vely, $J_{\rm DC}$ could be always enhanced in the ultra-quantum limit of strong magnetic fields, where the chiral LL becomes the only transport channel with all other 
nonchiral LLs pushed far away from the Fermi level. 
This na\"{i}ve expectation, however, does not hold since the chiral LL alone cannot induce any actual electronic transport, at least via elastic impurity scattering alone. 
In this case, nonzero net DC electric current can be generated with help of the broadening of Landau-Stark energy levels due to inelastic scattering processes.

Fig.~\ref{fig:EQO_chiral}~{\bf a} shows the behavior of $J_{\rm DC}$ as a function of $\Omega$ ranging from weak to strong electric fields in the ultra-quantum limit of strong magnetic fields, say, at $\omega_c=1$, where it is sufficient to consider only the chiral LL so long as $\Omega \lesssim 0.6$ (i.e., before the general Landau-Stark resonance comes into play). 
Particularly, in this limit, there are four distinct regimes of the chiral anomaly transport; (i) negative MR, (ii) non-Ohmic behavior, (iii) Esaki-Tsu peak, and (iv) EQO at weak ($\Omega \lesssim 0.01$), weak-to-intermediate ($0.01 \lesssim \Omega \lesssim 0.05$), intermediate ($\Omega=\Omega_{\rm ET} \simeq 0.05$), and strong ($\Omega \gtrsim 0.05$) electric fields, respectively.

First, at strong electric fields, the EQO occurs via the chiral resonance, which is distinguished from the previously described, general Landau-Stark resonance.
In the case of the chiral resonance, the DC electric current oscillates as a function of $1/\Omega$ with a constant period entirely independent of $\omega_c$, which is simply $\pi/4$ in units of $1/t_3$ in our minimal model for Weyl semimetals.
Fundamentally, the mechanism of the chiral resonance can be understood in terms of the wave function overlap between adjacent WSL eigenstates in the chiral LL, which oscillates asymptotically as a function of electric field.
See {\bf Methods} for details.

As $\Omega$ decreases, the EQO becomes less and less pronounced, finally merging into the Esaki-Tsu peak around $\Omega=\Omega_{\rm ET}$.
Fig.~\ref{fig:EQO_chiral}~{\bf b} shows that both Esaki-Tsu peak and subsequent EQO are closely correlated with the formation of well-separated WSL eigenstates.
Note that, marking the onset of negative differential conductivity, the Esaki-Tsu peak~\cite{Esaki_Tsu1970} has been routinely observed in semiconductor superlattices~\cite{Wacker2002}.

Fig.~\ref{fig:EQO_chiral}~{\bf c} shows that, at $\Omega \lesssim \Omega_{\rm ET}$, $J_{\rm DC}$ increases as a monotonic, but in general nonlinear function of $\Omega$, i.e., $J_{\rm DC} = \sigma \Omega +\sigma^\prime \Omega^2 +\sigma^{\prime\prime}\Omega^3 +\cdots$, where $\sigma$ denotes the usual linear Drude conductivity in the limit of weak electric fields, while $\sigma^\prime$ and $\sigma^{\prime\prime}$ are the two lowest-order coefficients of the non-Ohmic behavior. 
It is important to note that the non-Ohmic behavior is an inevitable crossover phenomenon connecting between the linear Drude conductivity and Esaki-Tsu peak. 
Considering that both Esaki-Tsu peak and subsequent EQO are closely correlated with the formation of well-separated WSL eigenstates, the non-Ohmic behavior can be regarded as a precursor to the EQO. 
In this context, the non-Ohmic behavior observed in BiSb alloys~\cite{Shin2017} suggests that the observation of EQO might actually be within the reach of experiments since, in our results, the strength of electric field necessary for the occurrence of EQO is only about $10\mbox{-}20$ times larger than that necessary for the non-Ohmic behavior.

Now, we would like to confirm if the linear Drude conductivity, $\sigma$, exhibits the expected behavior of negative MR. 
Specifically, in the ultra-quantum limit of strong magnetic fields, $\sigma$ is expected to increase as a linear function of magnetic field~\cite{Nielsen_Ninomiya1983,Son2013}.
Fig.~\ref{fig:EQO_chiral}~{\bf d} confirms that this is indeed exactly the case.  
Actually, Fig.~\ref{fig:EQO_chiral}~{\bf e} shows that $J_{\rm DC}$ increases as a whole with stronger magnetic fields. 
Finally, Fig.~\ref{fig:EQO_chiral}~{\bf f} shows the behavior of $J_{\rm DC}$ as a function of $\Omega$ for various given $V_{\rm imp}$, confirming that $\sigma$ decreases with stronger impurity scattering, as expected from the Drude behavior.

\noindent{\bf Discussion}

\noindent In this work, it is shown that the chiral charge pumping is essentially nothing but the Bloch oscillation. 
Both topologically and energetically protected in Weyl semimetals, the Bloch oscillation can be quantized to generate robust Landau-Stark eigenstates, eventually giving rise to the resonant oscillation of the DC electric current as a function of electric field.

Called the EQO, this resonant oscillation of the DC electric current can occur in Weyl semimetals via two different mechanisms. 
First, the EQO can occur via the resonance between various Landau-Stark eigenstates with different LL indices. 
Second, in the ultra-quantum limit of strong magnetic fields, the EQO can also occur via a form of the self-resonance within the chiral LL.
Particularly, in this limit, there are four distinct regimes of the chiral anomaly transport; (i) negative MR, (ii) non-Ohmic behavior, (iii) Esaki-Tsu peak, and (iv) EQO 
at weak, weak-to-intermediate, intermediate, and strong electric fields, respectively. 
It is important to note that both negative MR and non-Ohmic behavior~\cite{Shin2017} have been already observed in Weyl semimetals, providing experimental support for the occurrence of EQO in natural materials.

In broad perspective, understanding nonequilibrium steady states of matter is among the foremost frontiers in physics. 
Induced by strong electric fields, the EQO would be one of the most salient features of nonequilibrium steady states realized in condensed matter. 
Usually achieved in synthetic systems such as semiconductor superlattices and optical lattices, a prerequisite for the occurrence of EQO is the formation of robust WSL eigenstates. 
As emphasized in this work, the chiral anomaly can provide a unique environment for the formation of robust WSL eigenstates via the combination of strong-field phenomena with topological matter.
Interestingly, Weyl semimetals can be also synthetically generated by fabricating a layered structure of alternating topological and magnetic insulators~\cite{Burkov2011}.

Finally, there is a close analogy between the EQO studied in this work and the radiation-induced quantum oscillation observed in quantum Hall systems~\cite{Mani2002,Zudov2003}. 
It is interesting to mention that the radiation-induced quantum oscillation has been analyzed via both Keldysh nonequilibrium Green function method and tunneling formula~\cite{Shi2003,Durst2003,Park2004}, which are also two main theoretical tools in this work.

\noindent{\bf Methods}

\noindent{\bf Landau quantization in Weyl semimetals.}
We begin by writing the continuum limit of the model Hamiltonian in Eq.~\eqref{eq:H_k} within the $x$-$y$ plane, which can be obtained by replacing $\sin{k_i}$ by $k_i$ and $\cos{k_i}$ by $1-k_i^2/2$ for $i=x, y$, while maintaining the full $k_z$ dispersion.  
Specifically, the model Hamiltonian can be written in the continuum limit as follows:
\begin{equation}
H({\bf k}) = 2t_1 k_x \sigma_x - 2t_1 k_y \sigma_y + \left[2t_2 (k_x^2+k_y^2) + 2t_3\cos{k_z}\right] \sigma_z,
\label{eq:H_cont}
\end{equation}
where all momenta are denoted in units of corresponding inverse lattice constants.

With the application of magnetic fields in the $z$ direction, the model Hamiltonian is modified via minimal coupling, i.e., ${\bf k} \rightarrow \boldsymbol{\Pi}=-i\nabla+e{\bf A}_{\rm Landau}$ with ${\bf A}_{\rm Landau}=B(0,x,0)$ being the Landau-gauge vector potential. 
At this moment, let us assume that $B>0$.
The case of $B<0$ is to be considered separately below.
For $B>0$, the model Hamiltonian can be written as 
\begin{align}
H ({\bf k}) &=  
\frac{2\sqrt{2}t_1}{l_B} \left( b^{\dagger} \sigma_{+}  + b \sigma_{-} \right)
\nonumber\\
&+\left[ \omega_c \left(b^{\dagger} b + \frac{1}{2}\right) + 2 t_3 \cos{k_z} \right] \sigma_z,
\label{eq:H_minimal_coupling}
\end{align}
where the LL raising and lowering operators, $b^\dagger$ and $b$, are defined, respectively, as follows:
\begin{equation}
\left\{
\begin{array}{c}
b^\dagger \\
b
\end{array}
\right\}
= \frac{l_B}{\sqrt{2}} \left( \Pi_x \pm i \Pi_y  \right)  
\end{equation}
with $l_B=1/\sqrt{eB}$ being the magnetic length. 
Similarly, the pseudospin raising and lowering operators, $\sigma_{+}$ and $\sigma_{-}$, are defined, respectively, as follows:
\begin{equation}
\left\{
\begin{array}{c}
\sigma_{+} \\
\sigma_{-}
\end{array}
\right\}
= \frac{1}{2} \left( \sigma_x \pm i \sigma_y  \right).
\end{equation}
Note that the cyclotron frequency is given by $\omega_c=4 t_2 /l_B^2$, and $\Pi_z$ is replaced back to its eigenvalue, $k_z$.

The Hamiltonian in Eq.~\eqref{eq:H_minimal_coupling} can be block-diagonalized by using the convenient set of basis states, $\{ |\nu\rangle\otimes|\sigma\rangle \}$, which are composed of number eigenstates $|\nu\rangle$ (i.e., $b^\dagger b |\nu\rangle=\nu|\nu\rangle$) and the pseudospin up/down state $ |\sigma\rangle$ (i.e., $|\uparrow\;\rangle$ or $|\downarrow\;\rangle$). 
Now, by noting that
\begin{equation}
|\nu\rangle\otimes|\uparrow\;\rangle \xrightleftharpoons [b^{\dagger} \sigma_{+}]{b \sigma_{-}} | \nu-1\rangle\otimes|\downarrow\;\rangle, 
\end{equation}
one can obtain the block-diagonalized matrix form of the Hamiltonian as follows:
\begin{equation}
H_\nu(k_z) 
=\frac{\omega_c}{2} \mathbb{I}
+\left[
\begin{array}{cc}
2t_3\cos{k_z} +\nu\omega_c  & \frac{2\sqrt{2}t_1}{l_B}\sqrt{\nu}
\\
\frac{2\sqrt{2}t_1}{l_B}\sqrt{\nu}  &  -2t_3\cos{k_z} -\nu\omega_c
\end{array}
\right] ,
\label{eq:H_nu}
\end{equation}
which is defined in the Hilbert space spanned by two basis states, $|\nu\rangle\otimes|\uparrow\;\rangle$ and $|\nu-1\rangle\otimes|\downarrow\;\rangle$ with $\nu \geq 1$. 
Diagonalizing $H_\nu(k_z)$ generates the energy eigenvalues of nonchiral LLs as follows:
\begin{equation}
\epsilon_{\nu,\pm}(k_z)=\omega_c/2 \pm \sqrt{ (2t_3 \cos{k_z} + \nu\omega_c)^2 + 2\nu\omega_c t_1^2/t_2},
\label{eq:LL_energy_nonchiral2}
\end{equation}
which becomes identical to $\epsilon_n(k_z)$ in Eq.~\eqref{eq:LL_energy_nonchiral} after the LL index is defined as $n=\pm\nu$, and the zero-point energy $\omega_c/2$ is subtracted.

Meanwhile, the Hamiltonian is already fully diagonalized for $\nu=0$: 
\begin{equation}
H_0(k_z)=\omega_c/2+2t_3\cos{k_z},
\end{equation}
which is defined in the Hilbert space spanned by the single basis state, $|0\rangle\otimes|\uparrow\;\rangle$.
Being diagonal, $H_0(k_z)$ itself is the energy eigenvalue of the chiral LL, which equals to $\epsilon_{0}(k_z)$ after the subtraction of the zero-point energy.

It is important to note that the singleness of the chiral LL is a unique topological property of Weyl semimetals.
To appreciate the origin of this topological property, it is beneficial to consider what happens in the case of $B<0$. 
Actually, the model Hamiltonian can be written for the general sign of $B$ as follows:
\begin{align}
H ({\bf k}) &=  
\frac{2\sqrt{2}t_1}{l_B} \left( b^{\dagger} \sigma_{{\rm sgn}(B)}  + b \sigma_{-{\rm sgn}(B)} \right)
\nonumber\\
&+\left[ \omega_c \left(b^{\dagger} b + \frac{1}{2}\right) + 2 t_3 \cos{k_z} \right] \sigma_z,
\label{eq:H_minimal_coupling2}
\end{align}
where the LL raising and lowering operators are now generalized as follows:
\begin{equation}
\left\{
\begin{array}{c}
b^\dagger \\
b
\end{array}
\right\}
= \frac{l_B}{\sqrt{2}} \left[ \Pi_x \pm i {\rm sgn}(B) \Pi_y  \right] 
\end{equation}
with $l_B=1/\sqrt{e|B|}$.

After some algebra, one can show that the energy eigenvalues of nonchiral LLs are exactly the same as before regardless of the sign of $B$ except that the zero-point energy is now generalized as ${\rm sgn(B)}\omega_c/2$.

The situation is quite different for the chiral LL.
That is, unlike those of nochiral LLs, the energy eigenvalue of the chiral LL depends on the sign of $B$: $\epsilon_0(k_z)=2t_3{\rm sgn}(B)\cos{k_z}$. 
This sign dependence of the chiral LL is fundamentally due to the specific topological property of Weyl semimetals in our minimal model.
Namely, the 2D $k_z$ slices of the Brillouin zone form Chern or trivial insulators depending on whether $k_z$ is inside or outside the region between two Weyl nodes with opposite chiralities.

\noindent{\bf Noninteracting Green functions in the Floquet matrix form.}
\noindent Here, we discuss how to construct the noninteracting retarded and lesser Green functions in the Floquet matrix form. 
We begin by writing the noninteracting Hamiltonian in the temporal vector potential gauge as follows:
\begin{equation}
H= \sum_{n,k_z} \epsilon_n(k_z-\Omega t) c^\dagger_{n,k_z} c_{n,k_z},
\end{equation}
where $c^\dagger_{n,k_z}$ and $c_{n,k_z}$ are the creation and annihilation operators, respectively, for the $n$-th LL with $k_z$.

The noninteracting retarded Green function is defined as follows:
\begin{equation}
g_{n,k_z}^r(t,t^\prime) = -i\theta(t-t^\prime) \langle \{ c_{n,k_z}(t), c^\dagger_{n,k_z}(t^\prime) \} \rangle_0,
\label{eq:retarded_g}
\end{equation}
where $c_{n,k_z}(t) = U_{n,k_z}(t,t_0) c_{n,k_z}(t_0)$ with the unitary evolution operator, $U_{n,k_z}(t,t_0)$, given by 
\begin{equation}
U_{n,k_z}(t,t_0)={\rm exp}\left[-i\int_{t_0}^{t} d\tau \epsilon_n(k_z-\Omega \tau)\right]  ,
\label{eq:U}
\end{equation}
where $t_0$ is some arbitrary reference time.

Now, noting that the unitary evolution operator in Eq.~\eqref{eq:U} is essentially identical to the wave function of Landau-Stark eigenstates in Eq.~\eqref{eq:Landau-Stark_eigenstates}, $U_{n,k_z}(t,t_0)$ can be expressed in terms of $\phi_{nl}$ as follows:
\begin{equation}
U_{n,k_z}(t,t_0)=\frac{\phi_{nl}(k_z-\Omega t)}{\phi_{nl}(k_z-\Omega t_0)}  e^{-i\epsilon_{nl}(t-t_0)},
\label{eq:U2}
\end{equation}
where $l$ can be chosen arbitrarily. 
Then, plugging Eq.~\eqref{eq:U2} into the anticommutation part in Eq.~\eqref{eq:retarded_g} leads to the following result:
\begin{align}
&\langle \{ c_{n,k_z}(t), c^\dagger_{n,k_z}(t^\prime) \} \rangle_0 
\nonumber \\
&= \phi_{nl}(k_z-\Omega t) \phi_{nl}^*(k_z-\Omega t^\prime) e^{-i\epsilon_{nl}(t-t^\prime)} ,
\end{align}
where it is used that $\{ c_{n,k_z}(t_0), c^\dagger_{n,k_z}(t_0) \} =1$ and $\phi_{nl}(k_z-\Omega t_0)\phi^*_{nl}(k_z-\Omega t_0)=1$.

Next, by using the integral representation of the Heaviside step function,
\begin{equation}
-i\theta(t-t^\prime)=\int_{-\infty}^{\infty} \frac{d\varepsilon}{2\pi} \frac{e^{-i\varepsilon(t-t^\prime)}}{\varepsilon+i\eta} ,
\end{equation}
one can express $g_{n,k_z}^r(t,t^\prime)$ as follows:
\begin{align}
g_{n,k_z}^r(t,t^\prime) &= \phi_{nl}(k_z-\Omega t) \phi_{nl}^*(k_z-\Omega t^\prime) 
\nonumber \\
&\times \int_{-\infty}^{\infty} \frac{d\varepsilon}{2\pi} 
\frac{e^{-i\varepsilon(t-t^\prime)}}{\varepsilon-\epsilon_{nl}+i\eta} ,
\label{eq:retarded_g2}
\end{align}
which is obtained after an appropriate redefinition of the integration valuable.

Then, by using the Fourier transform of $\phi_{nl}$,
\begin{equation}
\phi_{nl}(k_z-\Omega t)= \sum_p e^{-ip(k_z-\Omega t)}\varphi_{nl}(p) ,
\end{equation} 
one can arrive at the final expression:
\begin{align}
g_{n,k_z}^r(t,t^\prime) &= \int_{-\Omega/2}^{\Omega/2} \frac{d\omega}{2\pi} \sum_{p,q}
e^{-i(\omega+p\Omega)t} e^{i(\omega+q\Omega)t^\prime}
\nonumber \\
&\times [g_n^r(k_z,\omega)]_{pq} ,
\label{eq:retarded_g3}
\end{align}
where
 \begin{equation}
[g_{n}^r(k_z,\omega)]_{pq}=
e^{ik_z(p-q)} \sum_{j} 
\varphi_{np}(j) {\cal G}_n^r(\omega+j\Omega) \varphi_{nq}^*(j) 
\label{eq:Floquet_g_retarded2}
\end{equation}
with ${\cal G}_n^r(\varepsilon=\omega+j\Omega)=1/(\varepsilon-\bar{\epsilon}_n+i\eta)$ being the reduced retarded Green function of Landau-Stark eigenstates. 
Note that the $l$ dependence completely disappears in the final expression due to the translational symmetry of Landau-Stark eigenstates, i.e., $\varphi_{np}(j)=\varphi_{n,p+l}(j+l)$ for arbitrary $l$.

The mathematical form of Eq.~\eqref{eq:retarded_g3} indicates that $[g_{n}^r(k_z,\omega)]_{pq}$ is nothing but the Fourier transform of $g_{n,k_z}^r(t,t^\prime)$.
Specifically, $[g_{n}^r(k_z,\omega)]_{pq}$ is the $(p,q)$-th element of the noninteracting retarded Green function in the Floquet matrix form~\cite{Tsuji2008}.

Based on this realization, it is instructive to compute the noninteracting local DOS, $\rho_{\rm loc}^{(0)}={\cal D} \sum_n \rho_{{\rm loc},n}^{(0)}$, where
\begin{align}
\rho_{{\rm loc},n}^{(0)}(\varepsilon=\omega+p\Omega) &= -\frac{1}{\pi} \int_{-\pi}^{\pi} \frac{dk_z}{2\pi} {\rm Im}[g_n^r(k_z,\omega)]_{pp}
\nonumber \\
&= 
\sum_{l} |\varphi_{nl}(0)|^2 \delta(\varepsilon-\epsilon_{nl}) ,
\label{eq:rho_loc0}
\end{align}
which shows that the local DOS is composed of discrete peaks at $\varepsilon=\epsilon_{nl}$ with their weights given by the corresponding Landau-Stark eigenstates at a given site, say, origin, $|\varphi_{nl}(0)|^2$.
Note that the broadening of Landau-Stark energy levels can be implemented by setting $\eta=\Gamma/2$ in ${\cal G}_n^r(\varepsilon)$ with $\Gamma$ being small, but finite, in which case the delta function is replaced by the Lorentzian:
\begin{equation}
\delta_{\Gamma}(\varepsilon-x) =-\frac{1}{\pi} {\rm Im}\frac{1}{\varepsilon-x+i\Gamma/2}
\label{eq:Lorentzian}
\end{equation}
with $\Gamma$ quantifying the broadening width.

Now, let us switch gears to the noninteracting lesser Green function.
Actually, Eq.~\eqref{eq:Floquet_g_retarded2} suggests a very natural mathematical expression for the noninteracting lesser Green function:
\begin{equation}
[g_{n}^<(k_z,\omega)]_{pq}=
e^{ik_z(p-q)} \sum_{j} 
\varphi_{np}(j) {\cal G}_n^<(\omega+j\Omega) \varphi_{nq}^*(j) 
\label{eq:Floquet_g_lesser2}
\end{equation}
where ${\cal G}_n^<(\varepsilon=\omega+j\Omega)$ is the reduced lesser Green function of Landau-Stark eigenstates.
As mentioned in the main text, ${\cal G}_n^<(\varepsilon)$ is obtained via the WSL-wise thermalization scheme~\cite{Lee2014}.
Specifically, each WSL eigenstate is individually thermalized according to the standard fluctuation-dissipation relation:
\begin{equation}
{\cal G}_n^<(\varepsilon) = [{\cal G}_n^a(\varepsilon)-{\cal G}_n^r(\varepsilon)] f_{\rm FD}(\varepsilon),
\end{equation}
where ${\cal G}_n^a(\varepsilon)={\cal G}_n^{r*}(\varepsilon)$ and $f_{\rm FD}(\varepsilon)=1/(e^{(\varepsilon-\mu)/k_B T}+1)$ is the usual Fermi-Dirac distribution function with the chemical potential $\mu$ set to be zero for half filling.

As done before, it is also instructive to compute the noninteracting local occupation number, $N_{\rm loc}^{(0)}={\cal D} \sum_n N_{{\rm loc},n}^{(0)}$, where
\begin{align}
N_{{\rm loc},n}^{(0)}(\varepsilon = \omega+p\Omega) &= \frac{1}{2\pi} \int_{-\pi}^{\pi} \frac{dk_z}{2\pi} {\rm Im}[g_n^<(k_z,\omega)]_{pp}
\nonumber \\
&= 
\sum_{l} \frac{|\varphi_{nl}(0)|^2 \delta(\varepsilon-\epsilon_{nl})}{e^{(\varepsilon-\mu-l\Omega)/k_B T}+1} ,
\label{eq:N_loc0}
\end{align}
which shows that each WSL eigenstate is indeed individually thermalized with its own effective chemical potential, $\mu_{\rm eff}(l) = \mu+l\Omega$.

\noindent{\bf DC electric current density from the full lesser Green function.}
The DC electric current can be computed from the full lesser Green function obtained as a converged solution of the Keldysh-Dyson self-consistency loop.

To begin with, whether DC or not,  the electric current density can be exactly expressed in terms of the full lesser Green function as follows:
\begin{equation}
J(t)=-{\cal D} e \int_{-\pi}^{\pi} \frac{dk_z}{2\pi} \sum_{n} \frac{\partial\epsilon_{n}(k_z-\Omega t)}{\partial k_z} 
\langle c_{n,k_z}^\dagger(t) c_{n,k_z}(t) \rangle ,
\label{eq:J_def}
\end{equation}
which can be understood as the sum of all contributions from each conduction mode, whose individual contribution is in turn given by the product between its group velocity and occupation number specified by the LL index $n$ and momentum $k_z$.
It is important to note that the above expression is in principle exact at arbitrary strengths of electric and magnetic fields. 
As shown below, eventually, the electric current becomes strictly DC in our situation.

First, by definition, the occupation number is equal to the equal-time full lesser Green function, which can be related to its Fourier transform as follows:
\begin{gather}
\langle c_{n,k_z}^\dagger(t) c_{n,k_z}(t) \rangle
= -iG_{n,k_z}^<(t,t)
\nonumber \\
= -i\int_{-\Omega/2}^{\Omega/2}\frac{d\omega}{2\pi}\sum_{p,q}e^{-i(p-q)\Omega t}[G_{n}^<(k_z,\omega)]_{pq} ,
\label{eq:equal_time_lesser_G}
\end{gather}
where $[G_{n}^<(k_z,\omega)]_{pq}$ is the $(p,q)$-th element of the full lesser Green function in the Floquet matrix form.

Next, the $k_z$ integration in Eq.~\eqref{eq:J_def} can be explicitly performed by using the constraint that 
\begin{equation}
[G_{n}^<(k_z,\omega)]_{pq}=e^{ik_z(p-q)}[G_{n}^<(k_z=0,\omega)]_{pq} ,
\label{eq:gauge_invariance}
\end{equation}
ensuring that $k_z$ always appears as the particular form of $k_z-\Omega t$. 
This constraint is a manifestation of the gauge invariance in our situation.

With help of Eq.~\eqref{eq:gauge_invariance}, one can then obtain the final expression for the electric current density:
\begin{equation}
J_{\rm DC} =
e {\cal D}
\int_{-\Omega/2}^{\Omega/2} \frac{d\omega}{2\pi}
\sum_{n,p,q}
p \bar{\epsilon}_{n}(p)
\left[ G_{n}^< (k_z=0,\omega) \right]_{p+q,q},
\label{eq:J_DC2}
\end{equation}
where $\bar{\epsilon}_{n}(p)=\int_{-\pi}^{\pi} \frac{dk_z}{2\pi}e^{i k_z p}\epsilon_{n}(k_z)$.
Note that the nominal time dependence in the initial expression disappears in the final expression after the $k_z$ integration.
Consequently, as mentioned before, the electric current becomes strictly DC.

\noindent{\bf Truncation of Floquet matrices.}
In the Floquet representation, Green functions are represented as infinite-dimensional Floquet matrices. 
For practical calculations, the dimension of Floquet matrices should be truncated with an appropriate cutoff limiting the range of Floquet indices.
In other words, we would like to represent retarded and lesser Green functions as finite-dimensional Floquet matrices, $[G^r(\omega)]_{pq}$ and $[G^<(\omega)]_{pq}$, respectively, with $p, q \in (0, \pm 1, \cdots, \pm {\cal L})$.
The cutoff ${\cal L}$ is determined via the following procedure.

To begin with, we first estimate the cutoff by requiring that the noninteracting local DOS is properly normalized for each individual LL.
Specifically, it can be said that the noninteracting local DOS  for the $n$-th LL is properly normalized if
\begin{equation}
\left| 1-\sum_{p=-{\cal L}_n}^{{\cal L}_n}\int_{-\Omega/2}^{\Omega/2}d\omega \rho_{{\rm loc},n}^{(0)}(\omega+p\Omega) \right| < \delta_{\rm tol},
\label{eq:cutoff_criterion}
\end{equation}
where $\rho_{{\rm loc},n}^{(0)}$ is given in Eq.~\eqref{eq:rho_loc0}, ${\cal L}_n$ is the cutoff for the $n$-th LL, and $\delta_{\rm tol}$ is a sufficiently small tolerance.
In this work, we set $\delta_{\rm tol}$ to be $10^{-8}$.

Finally, the overall cutoff ${\cal L}$ is chosen as the maximum of ${\cal L}_n$: ${\cal L}={\rm max}\{{\cal L}_n\}$.
As a general rule, the lower $\Omega$ becomes, the higher ${\cal L}$ is required. 
Roughly speaking, ${\cal L}$ is of the order of $1,000$ for $\Omega \lesssim 0.01$ while typically less than $100$ otherwise.

\noindent{\bf Mechanism of the chiral resonance.}
To understand the mechanism of the chiral resonance, we begin by rewriting the tunneling formula as follows:
\begin{equation}
J_{\rm tun} \propto \int_{-\infty}^{\infty} d\varepsilon \left[ \rho_{\rm loc}(\varepsilon+\Omega) N_{\rm loc}(\varepsilon) -\rho_{\rm loc}(\varepsilon) N_{\rm loc}(\varepsilon+\Omega) \right],
\label{eq:J_tun2}
\end{equation}
where $\rho_{\rm loc}$ and $N_{\rm loc}$ are the local DOS and occupation number, respectively.
Note that Eq.~\eqref{eq:J_tun2} is precisely identical to Eq.~\eqref{eq:J_tun} since $N_{\rm loc}=\rho_{\rm loc} f_{\rm loc}$ by definition.

Now, assuming that WSL eigenstates are well separated in the chiral LL, the local DOS can be accurately approximated as 
\begin{equation}
\rho_{\rm loc}(\varepsilon) \sim \sum_{l} A_l (\Omega)\delta_{\Gamma}(\varepsilon-l\Omega),
\end{equation}
where $A_l(\Omega)=|J_{l}(2t_3/\Omega)|^2$ and $\delta_{\Gamma}(\varepsilon)$ is the Lorentzian in Eq.~\eqref{eq:Lorentzian}.
Similarly, the local occupation number can be also accurately approximated as
\begin{equation}
N_{\rm loc}(\varepsilon) \sim \sum_{l} A_l(\Omega) \delta_{\Gamma}(\varepsilon-l\Omega) f_{\rm FD}(\varepsilon-l\Omega),
\end{equation}
which is obtained via the WSL-wise thermalization scheme as explained in Eq.~\eqref{eq:N_loc0}.

After some rearrangements, Eq.~\eqref{eq:J_tun2} can be rewritten as follows:
\begin{equation}
J_{\rm tun} \sim \sum_{m}
F_m(\Omega)
\left[ I_m(\Omega)-I_{-m}(\Omega) \right],
\end{equation}
where $F_m(\Omega)=\sum_l A_{l+1}(\Omega) A_{l+m}(\Omega)$
and
\begin{equation}
I_m(\Omega)
=
\int_{-\infty}^{\infty} d\varepsilon
\delta_{\Gamma}(\varepsilon+m\Omega) \delta_{\Gamma}(\varepsilon) f_{\rm FD}(\varepsilon)
\end{equation}
which is a monotonic function of $\Omega$ without any oscillatory behaviors.
This means that, if any, oscillatory behaviors should come from  $F_m(\Omega)$, which depends on the wave function form of WSL eigenstates in the chiral LL.

In the chiral LL, WSL eigenstates are described by the Bessel function, which can be approximated as
\begin{equation}
J_l (x) \approx  \sqrt{\frac{2}{\pi x}}  \cos{\left( x - \frac{l\pi}{2} - \frac{\pi}{4}  \right)}
\end{equation}
for $x \gg | l^2-1/4 |$.
After some algebra making use of this asymptotic behavior of the Bessel function, one can show that $F_m(\Omega)$ is an oscillatory function of $1/\Omega$ with the period of $\pi/4$ in units of $1/t_3$. 
In conclusion, the chiral resonance is due to the asymptotic, oscillatory behavior of WSL eigenstate wave functions in the chiral LL.



\noindent{\bf Acknowledgements}

\noindent The authors are grateful to Sutirtha Mukherjee and Jee Hoon Kim for various insightful discussions.
Also, the authors thank Center for Advanced Computation (CAC) at Korea Institute for Advanced Study (KIAS) for providing computing resources for this work.
This work is partially supported by the KIAS Individual Grants, PG032303 (KP) and PG071401 (KH), and the Army Research Office (ARO) under Grant No. W911NF2010013 and W911NF-16-1-0182 (WL).

\end{document}